\documentstyle[12pt]{article}
\newfont{\ffont}{msym10}                        
\newcommand{\beq}{\begin{equation}}             
\newcommand{\eeq}{\end{equation}}               
\newcommand{\bqry}{\begin{eqnarray}}            
\newcommand{\eqry}{\end{eqnarray}}              
\newcommand{\bqryn}{\begin{eqnarray*}}          
\newcommand{\eqryn}{\end{eqnarray*}}            
\newcommand{\NL}{\nonumber \\}                  
\newcommand{\preprint}[1]{\begin{table}[t]      
            \begin{flushright}                  
            \begin{large}{#1}\end{large}        
            \end{flushright}                    
            \end{table}}                        
\newcommand{\PD}[2]                             
    {\frac{\partial^{#2}}{\partial #1^{#2}}}    
\catcode`\@=11 \@addtoreset{equation}{section}  
\renewcommand{\theequation}                     
         {\arabic{section}.\arabic{equation}}   
\begin{document}
\preprint{TAUP-2149-94 \\  }
\title{On Relativistic Bose-Einstein Condensation}
\author{\\ L. Burakovsky,\thanks {Bitnet: BURAKOV@TAUNIVM.TAU.AC.IL.} \
L.P. Horwitz\thanks {Bitnet: HORWITZ@TAUNIVM.TAU.AC.IL. Also at
Department of Physics, Bar-Ilan University, Ramat-Gan, Israel  } \ \\  \\
School of Physics and Astronomy \\ Raymond and Beverly Sackler
Faculty of Exact Sciences \\ Tel-Aviv University, Tel-Aviv 69978, Israel
\\  \\ and \\  \\ W.C. Schieve\thanks{Bitnet: WCS@MAIL.UTEXAS.EDU.}\ \\
 \\ Ilya Prigogine Center \\ for Studies in Statistical Mechanics \\
University of Texas at Austin \\ Austin TX 78712, USA \\}
\date{ }
\maketitle
\begin{abstract}
We discuss the properties of an ideal relativistic gas of events
possessing Bose-Einstein statistics. We find that the mass spectrum of
such a system is bounded by $\mu \leq m\leq 2M/\mu _K,$ where $\mu $ is
the usual chemical potential, $M$ is an intrinsic dimensional scale
parameter for the motion of an event in space-time, and $\mu _K$ is an
additional mass potential of the ensemble. For the system including both
particles and antiparticles, with nonzero chemical potential $\mu ,$ the
mass spectrum is shown to be bounded by $|\mu |\leq m\leq 2M/\mu _K,$ and
a special type of high-temperature Bose-Einstein condensation can occur.
We study this Bose-Einstein condensation, and show that it corresponds to
a phase transition from the sector of continuous relativistic mass
distributions to a sector in which the boson mass distribution becomes
sharp at a definite mass $M/\mu _K.$ This phenomenon provides a mechanism
for the mass distribution of the particles to be sharp at some definite
value.
\end{abstract}

\bigskip
\section{Introduction}
There have been a number of papers in the past \cite{Al,BKM1,CR,BKM2},
which discuss the properties of an ideal relativistic Bose gas with
nonzero chemical potential $\mu .$ Particular attention has been given to
the behavior of the Bose-Einstein condensation and the nature of the
phase transition in $d$ space dimensions \cite{BKM2,Lan}. The basic work
was done many years ago by J\"{u}ttner \cite{Jut},
Glaser \cite{Gla}, and more recently by Landsberg and Dunning-Davies
\cite{LDD} and Nieto \cite{Nie}. These works were all done in the
framework of the usual on-shell relativistic statistical mechanics.

To describe an ideal Bose gas in the grand canonical ensemble, the
usual expression for the number of bosons $N$ in relativistic
statistical mechanics is
\beq
N=V\sum _{{\bf k}}n_k=V\sum _{{\bf k}}\frac{1}{e^{(E_k-\mu )/T}-1},
\eeq
where $V$ is the system's three-volume, $E_k=\sqrt{{\bf k}^2+m^2}$ and
$T$ is the absolute temperature (we use the system of units in which
$\hbar =c=k_B=1;$ we also use the metric $g^{\mu \nu }=(-,+,+,+)),$ and
one must require that $\mu \leq m$ in order to ensure a
positive-definite value for $n_k,$ the number of bosons with momentum
${\bf k}.$ Here $N$ is assumed to be a conserved quantity, so that it
makes sense to talk of a box of $N$ bosons. This can no longer be true
once $T\stackrel{>}{\sim }m$ [10]; at such temperatures quantum field
theory requires consideration of particle-antiparticle pair production.
If $\bar{N}$ is the number of antiparticles, then $N$ and $\bar{N}$ by
themselves are not conserved but $N-\bar{N}$ is. Therefore, the
high-temperature limit of (1.1) is not relevant in realistic physical
systems.

The introduction of antiparticles into the theory in a systematic way was
made by Haber and Weldon \cite {HW1,HW2}. They considered an ideal Bose
gas with a conserved quantum number (referred to as ``charge'') $Q,$
which corresponds to a quantum mechanical particle number operator
commuting with the Hamiltonian $\hat{H}.$ All thermodynamic quantities
may be then obtained from the grand partition function $Tr\;\{\exp \;[-(
\hat{H}-\mu \hat{Q})/T]\}$ considered as a function of $T,V,$ and $\mu $
\cite{Hua}. The formula for the conserved net charge, which replaces
(1.1), reads\footnote{One uses the standard recipe according to which all
additive thermodynamic quantities are reversed for antiparticles.} [10]
\beq
Q=V\sum _{{\bf k}}\left[\frac{1}{e^{(E_k-\mu )/T}-1}-
\frac{1}{e^{(E_k+\mu )/T}-1}\right].
\eeq
In such a formulation a boson-antiboson system is described by only one
chemical potential $\mu ;$ the sign of $\mu $ indicates whether particles
outnumber antiparticles or vice versa. The requirement that both $n_k$
and $\bar{n}_k$ be positive definite leads to the important relation
\beq
|\mu |\leq m.
\eeq
The sum over ${\bf k}$ in (1.2) can be replaced by an
integral, so that the charge density $\rho \equiv Q/V$ becomes
\beq
\rho =\frac{1}{2\pi ^2}\int _0^\infty k^2\;dk\left[
\frac{1}{e^{(E_k-\mu )/T}-1}-(\mu \rightarrow -\mu )\right],
\eeq
which is an implicit formula for $\mu $ as a function of $\rho $ and $T,$
and in the region $T>>m$ reduces to
\beq
\rho \cong \frac{\mu T^2}{3}.
\eeq
For $T$ above some critical temperature $T_c,$ one can always find a $\mu
$ ($|\mu |\leq m)$ such that (1.4) holds. Below $T_c$ no such $\mu $ can
be found, and (1.4) should be interpreted as the charge density of the
excited states: $\rho -\rho _0,$ where $\rho _0$ is the charge density of
the ground state [10] (with ${\bf k}=0;$ clearly, this state is given
with zero weight in the integral (1.4)). The critical temperature $T_c$
at which Bose-Einstein condensation occurs corresponds to $\mu =\pm m$
(depending on the sign of $\rho .)$ Thus, one sets $|\mu |=m$ in (1.4)
and obtains, via (1.5) (provided that $|\rho |>>m^3),$
\beq
T_c=\sqrt{\frac{3|\rho |}{m}}.
\eeq
Below $T_c,$ (1.4) is an equation for $\rho -\rho _0,$ so that the charge
density in the ground state is
\beq
\rho _0=\rho [1-(T/T_c)^2].
\eeq
It follows from Eq. (1.6) that any ideal Bose gas will condense at a
relativistic temperature ($T_c>>m),$ provided that $|\rho |>>m^3.$

Recently the analogous phenomenon has been studied in relativistic
quantum field theory \cite{HW2,Kap,BD,BBD}. For relativistic fields
Bose-Einstein condensation occurs at high temperatures and can be
interpreted in terms of a spontaneous symmetry breaking \cite{HW2}.

The extension and generalization of Bose-Einstein condensation to curved
space-times and space-times with boundaries has also been the subject of
much study. The non-relativistic Bose gas in the Einstein static universe
was treated in ref. \cite{Al}. The generalization to relativistic scalar
fields was given in refs. \cite{SP,PZ}. The extension to higher
dimensional spheres was given in ref. \cite{Shi}. Bose-Einstein
condensation on hyperbolic manifolds \cite{CV}, and in the Taub universe
\cite{Hua} has also been considered. More recently, by calculating the
high-temperature expansion of the thermodynamic potential when the
boundaries are present, Kirsten \cite{Kir} examined Bose-Einstein
condensation in certain cases. Later work of Toms \cite{Toms} showed how
to interpret Bose-Einstein condensation in terms of symmetry breaking, in
the manner of flat space-time calculations \cite{HW2,Kap}. The most
recent study by Lee {\it et al.} \cite{LKK} showed how interacting scalar
fields can be treated. Bose-Einstein condensation for self-interacting
complex scalar fields was considered in ref. \cite{KT0}.

In the present paper we consider a relativistic Bose gas within the
framework of a manifestly covariant relativistic statistical mechanics
\cite{HSP,HSS,BH1}. We shall review this framework briefly in the next
section. We obtain the expressions for characteristic thermodynamic
quantities and show that they coincide with those of the relativistic
on-shell theory, except for the value of the average energy (which
differs by a factor 2/3). We introduce antiparticles and discuss the
properties of Bose-Einstein condensation in such a particle-antiparticle
system. We show that it corresponds to a phase transition to a
high-temperature form of the usual on-shell relativistic kinetic theory.

\section{Relativistic $N$-body system}
In the framework of a manifestly covariant relativistic statistical
mechanics, the dynamical evolution of a system of $N$ particles, for the
classical case, is governed by equations of motion that are of the form
of Hamilton equations for the motion of $N$ $events$ which generate the
space-time trajectories (particle world lines) as functions of a
continuous Poincar\'{e}-invariant parameter $\tau ,$ called the
``historical time''\cite{Stu,HP}. These events are characterized by
their positions $q^\mu =(t,{\bf q})$ and energy-momenta $p^\mu =(
E,{\bf p})$ in an $8N$-dimensional phase-space. For the quantum case, the
system is characterized by the wave function $\psi _\tau (q_1,q_2,\ldots
,q_N)\in L ^2(R^{4N}),$ with the measure $d^4q_1d^4q_2\cdots d^4q_N\equiv
d^{4N}q,$ $(q_i\equiv q_i^\mu ;\;\;\mu =0,1,2,3;\;\;i=1,2,\ldots ,N),$
describing the distribution of events, which evolves with a generalized
Schr\"{o}dinger equation \cite{HP}. The collection of events (called
``concatenation'' \cite{AHL}) along each world line corresponds to a
{\it particle,} and hence, the evolution of the state of the $N$-event
system describes, {\it a posteriori,} the history in space and time of
an $N$-particle system.

For a system of $N$ interacting events (and hence, particles) one takes
\cite{HP}
\beq
K=\sum _i\frac{p_i^\mu p_{i\mu }}{2M}+V(q_1,q_2,\ldots ,q_N),
\eeq
where $M$ is a given fixed parameter (an intrinsic property of the
particles), with the dimension of mass, taken to be the same for all the
particles of the system. The Hamilton equations are
$$\frac{dq_i^\mu }{d\tau }=\frac{\partial K}{\partial p_{i\mu }}=\frac{p_
i^\mu }{M},$$
\beq
\frac{dp_i^\mu }{d\tau }=-\frac{\partial K}{\partial q_{i\mu }}=-\frac{
\partial V}{\partial q_{i\mu }}.
\eeq
In the quantum theory, the generalized Schr\"{o}dinger equation
\beq
i\frac{\partial }{\partial \tau }\psi _\tau (q_1,q_2,\ldots ,q_N)=K
\psi _\tau (q_1,q_2,\ldots ,q_N)
\eeq
describes the evolution of the $N$-body wave function
$\psi _\tau (q_1,q_2,\ldots ,q_N).$ To illustrate the meaning of this
wave function, consider the case of a single free event. In this case
(2.3) has the formal solution
\beq
\psi _\tau (q)=(e^{-iK_0\tau }\psi _0)(q)
\eeq
for the evolution of the free wave packet. Let us represent $\psi _\tau
(q)$ by its Fourier transform, in the energy-momentum space:
\beq
\psi _\tau (q)=\frac{1}{(2\pi )^2}\int d^4pe^{-i\frac{p^2}{2M}\tau }
e^{ip\cdot q}\psi _0(p),
\eeq
where $p^2\equiv p^\mu p_\mu ,$ $p\cdot q\equiv p^\mu q_\mu ,$ and $\psi
_0(p)$ corresponds to the initial state. Applying the Ehrenfest arguments
of stationary phase to obtain the principal contribution to $\psi _\tau
(q)$ for a wave packet at $p_c^\mu ,$ one finds ($p_c^\mu $ is the peak
value in the distribution $\psi _0(p))$
\beq
q_c^\mu \simeq \frac{p_c^\mu }{M}\tau ,
\eeq
consistent with the classical equations (2.2). Therefore,
the central peak of the wave packet moves along the classical
trajectory of an event, i.e., the classical world line.

In the case that $p_c^0=E_c<0,$ we see, as in Stueckelberg's classical
example \cite{Stu}, that $$\frac{dt_c}{d\tau }\simeq \frac{E_c}{M}<0.$$
It has been shown \cite{AHL} in the analysis of an evolution operator
with minimal electromagnetic interaction, of the form $$K=\frac{(p-eA(q)
)^2}{2M},$$ that the $CPT$-conjugate wave function is given by
\beq
\psi _\tau ^{CPT}(t,{\bf q})=\psi _\tau (-t,-{\bf q}),
\eeq
with $e\rightarrow -e.$ For the free wave packet, one has
\beq
\psi _\tau ^{CPT}(q)=\frac{1}{(2\pi )^2}\int d^4p
e^{-i\frac{p^2}{2M}\tau }e^{-ip\cdot q}\psi _0(p).
\eeq
The Ehrenfest motion in this case is $$q_c^\mu \simeq -\frac{p_c^\mu }{M}
\tau ;$$ if $E_c<0,$ we see that the motion of the event in the
$CPT$-conjugate state is in the positive direction of time, i.e.,
\beq
\frac{dt_c}{d\tau }\simeq -\frac{E_c}{M}=\frac{|E_c|}{M},
\eeq
and one obtains the representation of a positive energy generic event
with the opposite sign of charge, i.e., the antiparticle.

\subsection{Ideal relativistic Bose gas}
To describe an ideal gas of events obeying Bose-Einstein statistics in
the grand canonical ensemble, we use the expression for the number of
events found in \cite{HSP},
\beq
N=V^{(4)}\sum _{k^\mu }n_{k^\mu }=
V^{(4)}\sum _{k^\mu }\frac{1}{e^{(E-\mu -\mu _K\frac{m^2}{2M})/T}-1},
\eeq
where $V^{(4)}$ is the system's four-volume, $\mu _K$ the additional mass
potential \cite{HSP}, which we shall take, in order to simplify
subsequent consideration, to be a fixed parameter (which
determines an upper bound of the mass distribution in the ensemble we are
studying, as we shall see below), and $m^2\equiv -k^2=-k^\mu k_\mu .$
To ensure a positive-definite value for $n_{k^\mu },$ the number
of bosons with four-momentum $k^\mu ,$ we require that
\beq
m-\mu -\mu _K\frac{m^2}{2M}\geq 0.
\eeq
The discriminant for the l.h.s. of the inequality must be nonnegative,
i.e.,
\beq
\mu \leq \frac{M}{2\mu _K}.
\eeq
For such $\mu ,$ (2.11) has the solution
\beq
\frac{M}{\mu _K}\left( 1-\sqrt{1-\frac{2\mu \mu _K}{M}}\right) \leq m\leq
\frac{M}{\mu _K}\left( 1+\sqrt{1-\frac{2\mu \mu _K}{M}}\right) .
\eeq
For small $\mu \mu _K/M,$ the region (2.13) may be approximated by
\beq
\mu \leq m\leq \frac{2M}{\mu _K}.
\eeq
One sees that $\mu _K$ plays a fundamental role in determining an upper
bound of the mass spectrum, in addition to the usual lower bound $m\geq
\mu .$ In fact, small $\mu _K$ admits a very large range of off-shell
mass, and hence can be associated with the presence of strong
interactions \cite{MS}.

Replacing the sum over $k^\mu $ (2.10) by an integral, one obtains for
the density of events per unit space-time volume $n\equiv N/V^{(4)}$
\cite{ind},
\beq
n=\frac{1}{4\pi ^3}\int _{m_1}^{m_2}dm\int _{-\infty }^\infty d\beta
\;\frac{m^3\;\sinh ^2\beta }{
e^{(m\cosh \beta -\mu -\mu _K\frac{m^2}{2M})/T}-1},
\eeq
where $m_1$ and $m_2$ are defined in Eq. (2.13), and we have used the
parametrization \cite{HSS} $$\begin{array}{lcl}
p^0 & = & m\cosh \beta , \\
p^1 & = & m\sinh \beta \sin \theta \cos \phi , \\
p^2 & = & m\sinh \beta \sin \theta \sin \phi , \\
p^3 & = & m\sinh \beta \cos \theta ,
\end{array} $$ $$0\leq \theta <\pi ,\;\;\;0\leq \phi <2\pi ,\;\;\;-\infty
<\beta <\infty .$$

In this paper we shall restrict ourselves to the case of high temperature
alone:
\beq
T>>\frac{M}{\mu _K}.
\eeq
It is then possible (for simplicity here; see footnote 3 on Eq. (3.9)) to
neglect indistinguishability of bosons in the integrand\footnote{The
limits of integration remain as a vestige of the Bose-Einstein
distribution; they play a central role in the results, as we shall see
below.} and to rewrite (2.15) in the form
\beq
n=\frac{e^{\mu /T}}{4\pi ^3}\int _{m_1}^{m_2}m^3\;dm\int _{-\infty }^
\infty \sinh ^2\beta \;d\beta \;e^{-m\cosh \beta /T}e^{\mu _Km^2/2MT},
\eeq
which reduces, upon integrating out $\beta ,$ to \cite{BH1}
\beq
n=\frac{Te^{\mu /T}}{4\pi ^3}\int _{m_1}^{m_2}dm\;m^2K_1\left( \frac{m}{
T}\right) e^{\mu _Km^2/2MT},
\eeq
where $K_\nu (z)$ is the Bessel function of the third kind (imaginary
argument). Since $m\leq m_2\leq 2M/\mu _K,$
\beq
\frac{\mu _Km^2}{2MT}\leq \frac{\mu _K(2M/\mu _K)^2}{2MT}=\frac{2M}{T
\mu _K}<<1,
\eeq
in view of (2.16), and also
\beq
\frac{\mu }{T}\leq \frac{m}{T}\leq \frac{2M}{T\mu _K}<<1.
\eeq
Therefore, one can neglect the exponentials in Eq. (2.18), and for $K_1(
m/T)$ use the asymptotic formula \cite{AS}
\beq
K_\nu (z)\sim \frac{1}{2}\Gamma (\nu )\left( \frac{z}{2}\right) ^{-\nu },
\;\;\;z<<1.
\eeq
Thus, we obtain
\beq
n\cong \frac{T^2}{4\pi ^3}\int _{m_1}^{m_2}dm\;m=\frac{T^2}{2\pi ^3}
\left( \frac{M}{\mu _K}\right) ^2\sqrt{1-\frac{2\mu \mu _K}{M}}.
\eeq
 From this equation, one can identify the high-temperature
mass distribution for the system we are studying, $f(m)\sim m,$ so that
\beq
\langle m^\ell \rangle =\frac{\int _{m_1}^{m_2}dm\;m^{\ell +1}}
{\int _{m_1}^{m_2}dm\;m}=\frac{2}{\ell +2}\frac{m_2^{\ell +2}-m_1^{\ell
+2}}{m_2^2-m_1^2}.
\eeq
In particular,
\beq
\langle m\rangle =\frac{4}{3}\frac{M}{\mu _K}\left( 1-\frac{\mu \mu _K}{
2M}\right) ,
\eeq
\beq
\langle m^2\rangle =2\left( \frac{M}{\mu _K}\right )^2
\left( 1-\frac{\mu \mu _K}{M}\right) .
\eeq
Extracting the joint distribution for $\beta $ and $m$ from (2.17) in the
same way, let us also obtain the average values of the energy and the
energy squared for high $T.$ First,
\beq
\langle E\rangle \equiv \langle m\cosh \beta \rangle \cong \frac{\int
_{m_1}^{m_2}m^4dm\sinh ^2\beta \cosh \beta d\beta e^{-m\cosh \beta /T}}
{\int _{m_1}^{m_2}m^3dm\sinh ^2\beta d\beta e^{-m\cosh \beta /T}}.
\eeq
Integrating out $\beta ,$ one finds
\beq
\langle E\rangle \cong \frac{1}{4T}\frac{\int _{m_1}^{m_2}dm\;m^4[K_3(m/
T)-K_1(m/T)]}{\int _{m_1}^{m_2}dm\;m^2K_1(m/T)}.
\eeq
It is seen, with the help of (2.21), that it is possible to neglect $K_1$
in comparison with $K_3$ in the numerator of (2.27) and obtain, via
(2.21),
\beq
\langle E\rangle \cong \frac{1}{4T}\frac{\int _{m_1}^{m_2}dm\;m^4K_3(m/
T)}{\int _{m_1}^{m_2}dm\;m^2K_1(m/T)}\simeq 2T,
\eeq
in agreement with refs. \cite{HSP,HSS,BH1}. Similarly, one obtains
$$\langle E^2\rangle \equiv \langle m^2\cosh ^2\beta \rangle \cong \frac{
\int _{m_1}^{m_2}m^5dm\sinh ^2\beta \cosh ^2\beta d\beta e^{-m\cosh \beta
/T}}{\int _{m_1}^{m_2}m^3dm\sinh ^2\beta d\beta e^{-m\cosh \beta /T}}$$
\beq
=\frac{\int _{m_1}^{m_2}dm[m^4K_1(m/T)+3Tm^3K_2(m/T)]}{\int _{m_1}^
{m_2}dm\;m^2K_1(m/T)}\cong 3T\frac{\int _{m_1}^{m_2}dm\;m^3K_2(m/T)}{\int
_{m_1}^{m_2}dm\;m^2K_1(m/T)}\simeq 6T^2.
\eeq

Let us assume that the average $\langle p^\mu p^\nu \rangle $ has the
form
\beq
\langle p^\mu p^\nu \rangle =au^\mu u^\nu +bg^{\mu \nu },
\eeq
where $u^\mu =(1,{\bf 0})$ in the local rest frame. The values of $a$ and
$b$ can be then calculated as follows: for $\mu =\nu =0$ one has
$\langle (p^0)^2\rangle =a-b,$ while contraction of (2.30) with $g^{\mu
\nu }$ gives $-g^{\mu \nu }\langle p_\mu p_\nu \rangle =a-4b.$ The use of
the expressions (2.29) for $\langle (p^0)^2\rangle \equiv \langle E^2
\rangle ,$ and (2.25) for $-g^{\mu \nu }\langle p_\mu p_\nu \rangle
\equiv \langle m^2\rangle $ yields $$\left \{ \begin{array}{rcl}
a-b & = & 6T^2, \\
a-4b & = & 2(\frac{M}{\mu _K})^2\left( 1-\mu \mu _K/M\right) ,
\end{array} \right. $$ so that
\beq
a=8T^2-\frac{2}{3}\left( \frac{M}{\mu _K}\right) ^2
\left( 1-\frac{\mu \mu _K}{M}\right) ,
\eeq
\beq
b=2T^2-\frac{2}{3}\left( \frac{M}{\mu _K}\right) ^2
\left( 1-\frac{\mu \mu _K}{M}\right) .
\eeq
For $T>>M/\mu _K,$ it is possible to take $a\cong 8T^2,$ $b\cong 2T^2,$
and obtain, therefore,
\beq
\langle p^\mu p^\nu \rangle \cong 8T^2u^\mu u^\nu +2T^2g^{\mu \nu }.
\eeq

To find the expressions for
the pressure and energy density in our ensemble, we study the
particle energy-momentum tensor defined by the relation \cite{HSS}
\beq
T^{\mu \nu }(q)=\sum _i\int d\tau \frac{p_i^\mu p_i^\nu }{M/\mu _K}
\delta ^4(q-q_i(\tau )),
\eeq
in which $M/\mu _K$ is the value around which the mass of the bosons
making up the ensemble is distributed, i.e., it corresponds to the
limiting mass-shell value when the inequality (2.12) becomes equality.
Upon integrating over a small space-time volume $\triangle V$ and
taking the ensemble average, (2.34) reduces to \cite{HSS}
\beq
\langle T^{\mu \nu }\rangle =\frac{T_{\triangle V}}{M/\mu _K}n\langle
p^\mu p^\nu \rangle .
\eeq
In this formula $T_{\triangle V}$ is the average passage interval in
$\tau $ for the events which pass through the small (typical) four-volume
$\triangle V$ in the neighborhood of the $R^4$-point. The four-volume
$\triangle V$ is the smallest that can be considered a macrovolume in
representing the ensemble. Using the standard expression
\beq
\langle T^{\mu \nu }\rangle =(p+\rho )u^\mu u^\nu +pg^{\mu \nu },
\eeq
where $p$ and $\rho $ are the particle pressure and energy density,
respectively, we obtain
\beq
p=\frac{T_{\triangle V}}{\pi ^3}\frac{M}{\mu _K}\sqrt{1-\frac{2\mu
\mu _K}{M}}T^4,\;\;\;\rho =3p.
\eeq
To interpret these results we calculate the particle number density per
unit three-volume. The particle four-current is given by the
formula \cite{HSS}
\beq
J^\mu (q)=\sum _i\int d\tau \frac{p^\mu _i}{M/\mu _K}\delta ^4(q-q_i(
\tau )),
\eeq
which upon integrating over a small space-time volume and taking the
average reduces to
\beq
\langle J^\mu \rangle =\frac{T_{\triangle V}}{M/\mu _K}n\langle p^\mu
\rangle ;
\eeq
then
\beq
N_0\equiv \langle J^0\rangle =\frac{T_{\triangle V}}{M/\mu _K}n
\langle E\rangle ,
\eeq
so that
\beq
N_0=\frac{T_{\triangle V}}{\pi ^3}\frac{M}{\mu _K}\sqrt{1-\frac{2\mu
\mu _K}{M}}T^3,
\eeq
and we recover the ideal gas law
\beq
p=N_0T.
\eeq

Since, in view of (2.13), $$\frac{2M}{\mu _K}\sqrt{1-\frac{2\mu \mu _K}{
M}}=\triangle m$$ is a width of the mass distribution around the
value $M/\mu _K,$ Eqs. (2.37),(2.41) can be rewritten as
\bqry
p & = & \frac{T_{\triangle V}\triangle m}{2\pi ^3}T^4,\;\;\;
\rho \;=\;3p, \NL
N_0 & = & \frac{T_{\triangle V}\triangle m}{2\pi ^3}T^3.
\eqry
In ref. \cite{glim} we obtained a formula which relates the average
passage interval in $\tau ,$ $T_{\triangle V},$ with a width of the mass
distribution $\triangle m$ (in order that results of the off-shell theory
coincide with those of the standard on-shell theory in the
sharp-mass-shell limit):
\beq
T_{\triangle V}\triangle m=2\pi .
\eeq
One can understand this relation, up to a numerical factor, in terms of
the uncertainty principle (rigorous in the $L^2(R^4)$ quantum theory)
$\triangle E\cdot \triangle t\stackrel{>}{\sim }1/2.$ Since the time
interval for the particle to pass the volume $\triangle V$ (this smallest
macroscopic volume is bounded from below by the size of the wave packets)
$\triangle t\cong E/M \;\triangle \tau ,$ and the dispersion of $E$ due
to the mass distribution is $\triangle E\sim m\triangle m/E,$ one obtains
a lower bound for $T_{\triangle V}\triangle m$ of order unity.

Thus, with (2.44) holding, the formulas (2.43) finally reduce to
\bqry
p & = & \frac{T^4}{\pi ^2},\;\;\;\rho \;=\;3p, \\
N_0 & = & \frac{T^3}{\pi ^2},
\eqry
which are the standard expressions for high temperatures \cite{KT}. Thus,
the formulas for characteristic thermodynamic quantities and the equation
of state for a relativistic gas of off-shell events $coincide$ with those
of the relativistic gas of on-shell particles, except for the expression
for the average energy which takes the value $2T$ in the relativistic gas
of events, in contrast to $3T,$ as for the high-temperature limit of
the usual theory \cite{Pauli}. Experimental measurement of average energy
at high temperature can, therefore, affirm (or negate) the validity of
the off-shell theory. There seems to be no empirical evidence which
distinguishes between these results at the present time. The quantity
$\sigma =M_0c^2/k_BT,$ a parameter which distinguishes the relativistic
from the nonrelativistic regime (see, e.g., \cite{HSP}) is very large for
$M_0$ of the order of the pion mass, at ordinary temperatures; the
ultrarelativistic limit corresponding to $\sigma $ small becomes a
reasonable approximation for $T\stackrel{>}{\sim }10^{13}$ K.

\section{Introduction of antiparticles}
The introduction of (positive energy) antiparticles into the theory as
negative energy events in the $CPT$-conjugate state leads, by application
of the arguments of Haber and Weldon \cite{HW1}, or Actor \cite{Act}, to
a change in sign of $\mu $ in the distribution function for
antiparticles. We therefore write down the following relation which
represents the analog of the formula (1.2): \footnote{As for the
nonrelativistic theory, the ``free'' distribution functions describe
quasiparticles in a form which takes interactions into account entering
through the chemical potential. Since, by definition of a good
quasiparticle, it is not frequently emitted or absorbed. We therefore
consider the particles and antiparticles as two species. Since the
particle $number$ is determined by the derivative of the free energy with
respect to the chemical potential, $\mu $ must change sign for the
antiparticles \cite{HW1}. Similarly, the average mass (squared) is
obtained by the derivative with respect to $\mu _K$ \cite{HSP}; since
the mass (squared) of the antiparticle is positive, $\mu _K$ does not
change sign.}
\beq
N=V^{(4)}\sum _{k^\mu }\left[ \frac{1}{e^{(E-\mu -\mu _K\frac{m^2}{2M})
/T}-1}-\frac{1}{e^{(E+\mu -\mu _K\frac{m^2}{2M})/T}-1}\right] .
\eeq
We require that the both $n_{k^\mu }$'s in Eq. (3.1) be positive
definite. In this way we obtain the two quadratic inequalities,
\bqry
m-\mu -\mu _K\frac{m^2}{2M} & \geq  & 0, \NL
m+\mu -\mu _K\frac{m^2}{2M} & \geq  & 0,
\eqry
which give the following relation representing the nonnegativeness of
the corresponding discriminants:
\beq
-\frac{M}{2\mu _K}\leq \mu \leq \frac{M}{2\mu _K}.
\eeq
It then follows that we must consider the intersection of
the ranges of validity of the two inequalities (3.2). Indeed, if each
inequality is treated separately, there would be some values of $m$ for
which one and not another would be physically acceptable. One finds the
bounds of this intersection region by solving these inequalities, and
obtains\footnote{This is actually the solution of one of the inequalities
(the most restrictive), depending on the sign of $\mu .$}
\beq
\frac{M}{\mu _K}\left( 1-\sqrt{1-\frac{2|\mu |\mu _K}{M}}\right)\leq m
\leq \frac{M}{\mu _K}\left(1+\sqrt{1-\frac{2|\mu |\mu _K}{M}}\right) ,
\eeq
which for small $|\mu |\mu _K/M$ reduces, as in (2.14)
in the no-antiparticle case, to
\beq
|\mu |\leq m\leq \frac{2M}{\mu _K}.
\eeq

Replacing summation in (3.1) by integration, we now obtain a
formula for the number density:
\beq
n=\frac{1}{4\pi ^3}\int _{m_1}^{m_2}\!\!m^3dm\int _{-\infty }^\infty
\!\!\sinh ^2\beta d\beta \left[
\frac{1}{e^{(m\cosh \beta -\mu -\mu _K\frac{m^2}{2M})/T}-1}-
\frac{1}{e^{(m\cosh \beta +\mu -\mu _K\frac{m^2}{2M})/T}-1}\right] ,
\eeq
where $m_1$ and $m_2$ are defined in Eq. (3.4), which for large $T$
reduces, as above, to $$n=\frac{e^{\mu /T}-e^{-\mu /T}}{4\pi ^3}T\int
_{m_1}^{m_2}dm\;m^2K_1\left( \frac{m}{T}\right) e^{\mu _Km^2/2MT}.$$
Now, using the estimates (2.19),(2.20), and $\sinh (\mu /T)\cong \mu /T$
for $\mu /T<<1,$ we obtain (in place of (2.22))
\beq
n=\frac{1}{\pi ^3}\left( \frac{M}{\mu _K}\right) ^2
\sqrt{1-\frac{2|\mu |\mu _K}{M}}\mu T.
\eeq
In (3.7) $n$ is a conserved net event charge, the sign of $\mu $
indicating whether particles outnumber antiparticles or vice versa.
Similarly, one obtains
\bqryn
p & = & 2p(\mu ), \\
\rho  & = & 2\rho (\mu ),
\eqryn
where $p(\mu )$ and $\rho (\mu )$ are given by (2.37) with $\mu $
replaced by $|\mu |.$ On the other hand,
\beq
N_0=2\frac{T_{\triangle V}}{\pi ^3}\frac{M}{\mu _K}\sqrt{
1-\frac{2|\mu |\mu _K}{M}}\mu T^2,
\eeq
where the factor of $2\mu /T,$ as compared to (2.41), arises from the
difference between the factors $\exp (\pm \mu /T).$ One then obtain the
following expressions for the Bose gas including both particles and
antiparticles\footnote{If we did not neglect indistinguishability of
bosons at high temperature, we would obtain, instead of (2.46)
\cite{glim}, $N_0=\frac{T^3}{\pi ^2}Li_3(e^{\mu /T}),$ where $Li_\nu (z)
\equiv \sum _{s=1}^\infty z^s/s^\nu $ is the polylogarithm \cite{Prud},
so that, for the system including both particles and antiparticles,
$N_0=\frac{T^3}{\pi ^2}[Li_3(e^{\mu /T})-Li_3(e^{-\mu /T})].$ It then
follows from the properties of the polylogarithms \cite{Prud} that, for
$x\equiv |\mu |/T<<1,$ $Li_3(e^x)-Li_3(e^{-x})\cong \frac{\pi ^2}{3}x,$
so that, we would obtain, instead of (3.10), $N_0=\mu T^2/3,$ which
coincides with Haber and Weldon's Eq. (1.5).}:
\bqry
p & = & \frac{2T^4}{\pi ^2},\;\;\;\rho \;=\;3p, \\
N_0 & = & \frac{2T^2}{\pi ^2}\mu .
\eqry

\subsection{Relativistic Bose-Einstein condensation}
Let us show that the expression for $N_0$ (2.40) coincides with the
thermodynamic definition
\beq
N_0=\frac{N}{V},
\eeq
where $N$ is the number of bosons in a three-dimensional box of volume
$V.$ Since the event number density $n$ is, by definition,
$$n=\frac{N}{V^{(4)}}=\frac{N}{V\triangle t},$$ where $\triangle t$ is
the (average) extent of the ensemble along the $q^0$-axis (as in our
discussion after (2.44)), one has
\beq
N_0=n\triangle t.
\eeq
The equation of motion (2.2) for $q^0$ (with $M/\mu _K,$ the central
value of the mass distribution, instead of $M,$
which corresponds to a change of scale parameter in the expression (2.1)
for the generalized Hamiltonian $K),$ $$\frac{dq_i^0}{d\tau }=\frac{p_i^0
}{M/\mu _K},$$ upon averaging over the whole ensemble, reduces to
\beq
\frac{\triangle t}{T_{\triangle V}}=\frac{\langle E\rangle }{M/\mu _K},
\eeq
where $T_{\triangle V}$ is the average passage interval in $\tau $ used
in previous consideration. Then, in view of (3.12),(3.13), one obtains
the equation (2.40).

Since in the particle-antiparticle case, $N_{{\rm rel}}\equiv N-\bar{N},$
where $N$ and $\bar{N}$ are the numbers of particles and antiparticles,
respectively, is a conserved quantity, according to the arguments of
Haber and Weldon \cite{HW1} pointed out in
Section I, $N_0=N_{{\rm rel}}/V$ is also a conserved quantity, so that
it makes sense to talk of $|N_{{\rm rel}}|$ bosons in a spatial box of
the volume $V.$ Therefore, in Eq. (3.10) $N_0$ is a conserved quantity,
so that, the dependence of $\mu $ on temperature is defined by (we
assume that $N_0$ is continuous at the phase transition)
\beq
\mu =\frac{\pi ^2N_0}{2T^2}.
\eeq

For $T$ above some critical temperature, one can always find $\mu $
satisfying (3.3) such that the relation (3.14) holds; no such $\mu $
can be found for $T$ below the critical temperature. The value of the
critical temperature is defined by putting $|\mu |=M/2\mu _K$ in (3.14):
\beq
T_c=\pi \sqrt{\frac{|N_0|}{M/\mu _K}}.
\eeq
For $|\mu |=M/2\mu _K,$ the width of the mass distribution is zero, in
view of (3.4), and hence the ensemble approaches a distribution sharply
peaked at the mass-shell value $M/\mu _K.$ The fluctuations $\delta m=
\sqrt{\langle m^2\rangle -\langle m\rangle ^2}$ also vanish. Indeed, as
follows from (2.24),(2.25) with $\mu $ replaced by $|\mu |,$ and
(3.14),(3.15),
\beq
\delta m=\frac{M}{3\mu _K}\sqrt{2-\left( \frac{T_c}{T}\right) ^2-
\left( \frac{T_c}{T}\right) ^4},
\eeq
so that, at $T=T_c,$ $\delta m=0.$ It follows from (3.16) that for $T$
in the vicinity of $T_c$ $(T\geq T_c),$
\beq
\delta m\simeq \frac{M}{3\mu _K}\sqrt{\frac{6}{T_c}}\sqrt{T-T_c}.
\eeq

We note that Eqs. (2.45),(2.46) do not contain explicit dependence on the
chemical potential, and hence no phase transition is induced. In fact, at
lower temperature (or small $\mu _K)$ one or the other of the particle or
antiparticle distribution dominates, and one returns to the case of the
high-temperature strongly interacting gas \cite{BHS1}. The remaining
phase transition is the usual low-temperature Bose-Einstein condensation
discussed in the textbooks.

One sees, with the help of (3.4), that the expression for $n$ (3.7) can
be rewritten as
\beq
n=\frac{1}{2\pi ^3}\frac{M}{\mu _K}\triangle m\;\mu T;
\eeq
since at $T=T_c,$ $\triangle m=0,$ it follows that $n=0$ at all
temperatures below $T_c.$ Therefore, the behaviour of an
ultrarelativistic Bose gas including both particles and antiparticles,
which is governed by the relation (3.14), can be thought of as a special
type of Bose-Einstein condensation to a ground state with $p^\mu p_\mu
=-(M/\mu _K)^2$ (this ground state occurs with zero weight in the
integral (3.6)). In such a formulation, every state with temperature
$T>T_c,$ given by Eq. (3.6), should be considered as an {\it off-shell}
excitation of the on-shell ground state. At $T=T_c,$ all such excitations
freeze out and the distribution becomes strongly peaked at a definite
mass, i.e., the system undergoes a phase transition to the on-shell
sector. Note that, for $n=0,$ Eq. (3.12) gives $\triangle t=0.$ Then,
since $\langle E\rangle \sim T,$ one obtains from (3.13) that $T_{
\triangle V}=0$ (this relation can be also obtained from (2.44) for
$\triangle m=0),$ which means that all the events become particles.

Since in both (off-shell and on-shell) phases the temperature
dependence of pressure and energy density are the same, and the velocity
of sound, $c^2\equiv dp/d\rho ,$ is also the same, this phase transition
is a second order phase transition. As the distribution function enters
the on-shell phase at $T=T_c,$ the underlying off-shell theory describes
fluctuations around the sharp mean mass. This phenomenon provides a
mechanism, based on equilibrium statistical mechanics, for understanding
how the general off-shell theory is constrained to the neighborhood of a
sharp universal mass shell for each particle type. At temperatures below
$T_c,$ the results of the theory for the main thermodynamic quantities
coincide with those of the usual on-shell theories.

In order that our considerations be valid, there must hold the relation
$T_c>>M/\mu _K,$ which reduces, via (3.15), to
\beq
\Big| N_0\Big| >>\frac{1}{\pi ^2}\left( \frac{M}{\mu _K}\right) ^3.
\eeq
For $M/\mu _K\sim m_\pi \simeq 140$ MeV, this inequality yields $N_0>>
3\cdot 10^5\;{\rm MeV}^3.$ Taking $N_0\sim 5\cdot 10^6\;{\rm MeV}^3,$
which corresponds to temperature $\sim 350$ MeV, in view of (2.46),
one gets $T_c\sim 550\;{\rm MeV}\simeq 4m_\pi .$

If $\mu _K$ is very small, it is difficult to satisfy (3.19) and the
possibility of such a phase transition may disappear. This case
corresponds, as noted above, to that of strong interactions and is
discussed in succeeding paper \cite{BHS1}.

\section{Concluding remarks}
We have considered the ideal relativistic Bose gas within the framework
of a manifestly covariant relativistic statistical mechanics, taking
account of antiparticles. We have shown that in such a
particle-antiparticle system at some critical temperature $T_c$ a
special type of relativistic Bose-Einstein condensation sets in, which
corresponds to phase transition from the sector of relativistic mass
distributions to a sector in which the boson mass distribution peaks at a
definite mass. The results which can be computed from the latter
coincide with those obtained in a high-temperature limit of the usual
on-shell relativistic theory.

The relativistic Bose-Einstein condensation in particle-antiparticle
system considered in the present paper can represent, along with the
Galilean limit $c\rightarrow \infty $ \cite{glim}, a possible mechanism
of acquiring a given sharp mass by the particles of the system, as a
phase transition between the corresponding sectors of the theory. Since
this phase transition can occur at an ultrarelativistic temperature, it
might be relevant to cosmological models. The relativistic Bose-Einstein
condensation considered in the present paper may also have properties
which could be useful in the study of relativistic boson stars
\cite{Jet}. These and the other aspects of the theory are now under
further investigation.


\bigskip
\bigskip
\begin{thebibliography}{9}
\bibitem{Al} M.B. Al'taie, J. Phys. A {\bf 11} (1978) 1603
\bibitem{BKM1} R. Beckmann, F. Karsch and D.E. Miller, Phys. Rev. Lett.
{\bf 43} (1979) 1277
\bibitem{CR} C.A. Arag\~{a}o de Carvalho and S. Goulart Rosa, Jr.,
J. Phys. A {\bf 13} (1980) 989
\bibitem{BKM2} R. Beckmann, F. Karsch and D.E. Miller, Phys. Rev. A {\bf
25} (1982) 561
\bibitem{Lan} P.T. Landsberg, in {\it Statistical Mechanics of Quarks and
Hadrons,} ed. H. Satz, (North-Holland, Amsterdam, 1981)
\bibitem{Jut} F. J\"{u}ttner, Z. Phys. {\bf 47} (1928) 542
\bibitem{Gla} W. Glaser, Z. Phys. {\bf 94} (1935) 677
\bibitem{LDD} P.T. Landsberg and J. Dunning-Davies, Phys. Rev. A
{\bf 138} (1965) 1049
\bibitem{Nie} M.M. Nieto, Lett. Nuovo Cimento {\bf 1} (1969) 677; J.
Math. Phys. {\bf 11} (1970) 1346
\bibitem{HW1} H.E. Haber and H.A. Weldon, Phys. Rev. Lett.
{\bf 46} (1981) 1497
\bibitem{HW2} H.E. Haber and H.A. Weldon, Phys. Rev. D {\bf 25} (1982)
502
\bibitem{Hua} K. Huang, {\it Statistical Mechanics,}
(Wiley, New York, 1963)
\bibitem{Kap} J.I. Kapusta, Phys. Rev. D {\bf 24} (1981) 426
\bibitem{BD} J. Bernstein and S. Dodelson, Phys. Rev. Lett. {\bf 66}
(1991) 683
\bibitem{BBD} K.M. Benson, J. Bernstein and S. Dodelson, Phys. Rev. D
{\bf 44} (1991) 2480
\bibitem{SP} S. Singh and R.K. Pathria, J. Phys. A {\bf 17} (1984) 2983
\bibitem{PZ} L. Parker and Y. Zhang, Phys. Rev. D {\bf 44} (1991) 2421
\bibitem{Shi} K. Shiraishi, Prog. Theor. Phys. {\bf 77} (1987) 975
\bibitem{CV} G. Cognola and L. Vanzo, Phys. Rev. D {\bf 47} (1993) 4575
\bibitem{Hu} W. Huang, J. Math. Phys. {\bf 35} (1994) 3594
\bibitem{Kir} K. Kirsten, Class. Quant. Grav. {\bf 8} (1991) 2239;
J. Phys. A {\bf 24} (1991) 3281
\bibitem{Toms} D.J. Toms, Phys. Rev. Lett. {\bf 8} (1992) 1152; Phys.
Rev. D {\bf 47} (1993) 2483
\bibitem{LKK} M.-H. Lee, H.-C. Kim and J.K. Kim, Bose-Einstein
Condensation for a Self-Interacting Theory in Curved Spacetimes, Preprint
KAIST-CHEP-93/M4
\bibitem{KT0} K. Kirsten and D.J. Toms, Bose-Einstein Condensation for
Interacting Scalar Fields in Curved Spacetime, Preprint UB-ECM-PF 94/37
\bibitem{HSP} L.P. Horwitz, W.C. Schieve and C. Piron,
Ann. Phys. (N.Y.)  {\bf 137} (1981) 306
\bibitem{HSS} L.P. Horwitz, S. Shashoua and W.C. Schieve,
Physica A  {\bf 161} (1989) 300
\bibitem{BH1} L. Burakovsky and L.P. Horwitz, Physica A, {\bf 201} (1993)
666
\bibitem{Stu} E.C.G. Stueckelberg, Helv. Phys. Acta {\bf 14} (1941)
372, 588; {\bf 15} (1942) 23
\bibitem{HP} L.P. Horwitz and C. Piron, Helv. Phys. Acta {\bf 46} (1973)
316
\bibitem{AHL} R. Arshansky, L.P. Horwitz and Y. Lavie, Found. Phys.
{\bf 13} (1983) 1167
\bibitem{MS} D.E. Miller and E. Suhonen, Phys. Rev. D {\bf 26} (1982)
2944
\bibitem{ind} L. Burakovsky and L.P. Horwitz, Found. Phys. {\bf 25}
(1995) 785
\bibitem{AS} M. Abramowitz and I.A. Stegun, {\it Handbook of Mathematical
Functions,} (Dover, New York, 1972), p. 375
\bibitem{glim} L. Burakovsky and L.P. Horwitz,
J. Phys. A {\bf 27} (1994) 4725
\bibitem{KT} See, for example, E.W. Kolb and M.S. Turner, {\it The Early
Universe,} (Addison-Wesley, Redwood, CA, 1990), p. 62
\bibitem{Pauli} W. Pauli, {\it Theory of Relativity,} (Pergamon, Oxford,
1958)
\bibitem{Act} A. Actor, Nucl. Phys. B {\bf 256} (1986) 689
\bibitem{Prud} A.P. Prudnikov {\it et al.,} {\it Integrals and Series,}
(Gordon and Breach, New York, 1980), Vol. 3, p. 762, Appendix II.5
\bibitem{BHS1} L. Burakovsky, L.P. Horwitz and W.C. Schieve, On the
Relativistic Statistical Mechanics of Strongly Interacting Matter,
Preprint TAUP-2231-95
\bibitem{Jet} Ph. Jetzer, Phys. Repts. {\bf 220} (1992) 163
\end{thebibliography}
\end{document}